\documentclass[manuscript,screen,nonacm]{acmart}

\AtBeginDocument{%
  \providecommand\BibTeX{{%
    \normalfont B\kern-0.5em{\scshape i\kern-0.25em b}\kern-0.8em\TeX}}}

\usepackage{color,soul}
\usepackage{float}
\restylefloat{table}
\usepackage[section]{placeins}
\usepackage{etoolbox}

\setcopyright{none}

\usepackage{blindtext}
\usepackage{xpatch}

\makeatletter
\xpatchcmd{\ps@firstpagestyle}{Manuscript submitted to ACM}{}{\typeout{First patch succeeded}}{\typeout{first patch failed}}
\xpatchcmd{\ps@standardpagestyle}{Manuscript submitted to ACM}{}{\typeout{Second patch succeeded}}{\typeout{Second patch failed}}    \@ACM@manuscriptfalse
\makeatother

\renewcommand\footnotetextcopyrightpermission[1]{} 
\setcopyright{none}
\begin{document}

\title{Auditing Work: Exploring the New York City algorithmic bias audit regime  }

\author{Lara Groves}
\email{lgroves@adalovelaceinstitute.org}
\affiliation{%
  \institution{Ada Lovelace Institute}
  \country{United Kingdom} 
}

\author{Jacob Metcalf}
\email{jake.metcalf@datasociety.net}
\affiliation{%
  \institution{Data \& Society Research Institute}
  \country{USA} 
}

\author{Alayna Kennedy}
\email{alayna.a.kennedy@gmail.com}
\affiliation{%
  \institution{Independent Researcher}
  \country{USA} 
}

\author{Briana Vecchione}
\email{briana@datasociety.net}
\affiliation{%
  \institution{Data \& Society Research Institute}
  \country{USA} 
}

\author{Andrew Strait}
\email{AStrait@adalovelaceinstitute.org}
\affiliation{%
  \institution{Ada Lovelace Institute}
  \country{United Kingdom} 
}

\renewcommand{\shortauthors}{Groves, et al.}

\begin{abstract}
In July 2023, New York City (NYC) implemented the first 
attempt to create an algorithm auditing regime for commercial machine-learning systems. Local Law 144 (LL 144), requires NYC-based 
employers using automated employment decision-making tools (AEDTs) in hiring to be subject to annual bias audits by an independent 
auditor. In this paper, we analyse what lessons can be learned from LL 144 for other national attempts to create algorithm auditing regimes. 
Using qualitative interviews with 16 experts and practitioners working within the regime, we find LL 144 has failed to create an effective 
auditing regime:  the law fails to clearly define key aspects like AEDTs and what constitutes an independent auditor, leaving auditors, 
vendors who create AEDTs, and companies using AEDTs to define the law’s practical implementation in ways that failed to protect job 
applicants. Several factors contribute to this: first, the law was premised on a faulty transparency-driven theory of change that fails to stop 
biased AEDTs from being used by employers. Second, industry lobbying led to the definition of what constitutes an AEDT being narrowed 
to the point where most companies considered their tools exempt. Third, we find auditors face enormous practical and cultural challenges 
gaining  access  to  data  from  employers  and  vendors  building  these  tools.  Fourth,  we  find  wide  disagreement  over  what  constitutes  a 
legitimate auditor and identify four different kinds of ‘auditor roles’ that serve different functions and offer different kinds of services.  
We conclude with four recommendations for policymakers seeking to create similar bias auditing regimes that use clearer definitions and 
metrics and more accountability. By exploring LL 144 through the lens of auditors, our paper advances the evidence base around audit as 
an accountability mechanism, and can provide guidance for policymakers seeking to create similar regimes.
\end{abstract}



\maketitle

\section{INTRODUCTION}
In 2023, New York City (NYC) became the first jurisdiction to implement a law that mandates independent algorithmic 
bias audits for commercial companies and city agencies, specifically focusing on automated employment decision-making 
tools (AEDTs) in hiring and promotion. Under this law—known as Local Law 144 (LL 144)—all NYC-based employers 
using AEDTs are obligated to hire a third-party independent auditor to conduct annual ‘bias audits’ and post the resultant 
audit reports on their website. Additionally, the law requires that those employers provide job-seekers with a transparency 
notice regarding their use of AEDTs and the right to opt-out of analysis by the algorithmic system in favour of a human 
decision process.   

LL 144 is the first law to create a third-party algorithm audit regime for AI and machine-learning systems. Algorithm 
audits are an emerging method for assessing an AI system for a particular kind of legal or ethical risk. They aim to create 
greater accountability for AI system developers \cite{goodman2022algorithmic} or inscribe normative standards\cite{leslie2019understanding}. Audits can encompass a range 
of different practices and assess for different risks \cite{Brennan_AI_assurance}. This includes technical audits of a system’s inputs and outputs to determine if the system performs differently for different user groups \cite{raji2019actionable}, auditing whether a system complies with local 
regulation  or  internal  standards of  development,  and  sociotechnical  audits  that  assess  how  a  system  is  impacting  wider 
societal processes and contexts \cite{radiya2023sociotechnical}. Auditing methodologies stem primarily from the fields of computer science and data 
science, and have largely focused on audits assessing issues of bias in AI systems\cite{buolamwini2018gender}. Previous research has highlighted 
that audits conducted by an independent third party tend to be of higher quality than audits conducted by an internal team 
or a contracted second-party assessor \cite{bennett2013customer, deis1992determinants, tepalagul2015auditor}. In other sectors, like finance or environmental studies, third-party auditing 
regimes require underlying governance mechanisms—such as transparency requirements to publish the audit, standards of 
auditor practice, and an oversight body to adjudicate instances of malpractice. These create a functional auditing ecosystem 
in which multiple actors can evaluate, test and audit AI systems before, during and after they are deployed\cite{raji2022outsider}.

While LL 144 is the first law to create an algorithm auditing regime, there is some history of regulators and government 
agencies using algorithm audits to assess bias in AI systems. Some national regulators in regions like Australia ACCC \cite{ausCommission2020} 
and the Netherlands \cite{Rekenkamer2022} have the power to audit algorithms to assess their legality or compliance with national regulation 
or  law,  including  issues  of  algorithmic  bias.  An  increasing  number  of  global  AI  governance  policy  proposals  are  now 
seeking to establish auditing regimes as part of a wider AI governance process. The European Union’s Digital Services 
Act EC \cite{EU_DSA} which comes into force in February 2024, requires ‘very large online platforms’ to conduct regular audits of 
their compliance with the law, which may often include audit practices similar to LL 144. Similar auditing requirements 
are also being proposed in legislation in the United Kingdom \cite{UKonline2022} and the United States Congress \cite{Clarke2022}. The US state of 
Colorado also passed a statute to prevent bias in the insurance industry through commercial algorithmic systems shortly 
after LL 144, which went into effect November, 2023, although that does not require independent audits \cite{coloradoInsurance2023,PG2023}.  

The theory of change animating LL 144 is an assumption common in policymaking: by incentivising and supporting a 
market for assessment and transparency documentation, assessors will be able to propagate best practices and protect civil 
rights even without strict regulatory limits on the design of technical systems. Additionally, by providing the public—in 
this case job-seekers—with insight into the algorithmic systems making impactful decisions about their lives, the public 
will make better informed decisions and push the vendors building these tools and the companies  using them to create 
fair(er) systems. In this study, we explore the experience of those conducting independent audits to illustrate whether the 
particular accountability structures imposed by this first-of-its-kind law will achieve those goals. We centre the experience 
of auditors in this study because they are the node who must interpret the needs and interests of employers,  developers, 
regulators, and the jobseekers the law is meant to protect, all while providing a financially viable service.  
Using  qualitative  interviews  with  16  experts  and  practitioners  working  within  the  LL  144  bias  audit  regime,  our  paper 
seeks to answer three research questions (RQs):  

\begin{itemize}
    \item \textbf{RQ1:} What are the practical components of a bias audit in this context? 
    \item \textbf{RQ2:} What are the relational dynamics and incentives that make for an effective bias auditing regime?  
    \item \textbf{RQ3:} What are the experiences of auditors, and how can they inform wider policy and practice?  
\end{itemize}
 
The core finding of our study is that the work of algorithm auditing in this regime is largely about managing the relational 
dynamics between stakeholders established by the accountability structures in the law. This paper offers novel empirical 
evidence into an emerging bias audit regime with the aim of informing wider policy and practice around algorithm audits 
and auditors at the local, national, and international level. To our knowledge, this is the sole study exploring the LL 144 
regime through the lens of auditors and one of the first attempts to use a case study of a bias auditing regime to discern 
lessons for other attempts. By developing the evidence base around audits, we can draw stronger conclusions about auditing as  a  potential  accountability  mechanism  and  contribute  to  an  ecosystem  of  safe  and  ethical  AI  systems  in  high-stakes 
decision-making domains like recruitment.

\section{METHODS}
\subsection{Expert interviews} 
Recruitment for interviews followed a combination of direct recruitment and snowball recruitment: we contacted known 
auditors in the LL 144 via the publicly available audits, through contacting auditors in our own personal networks, and 
from recommendations from our interviewees. We sent a total of 42 invitations The interviews were led by the lead author, 
with support and contributions from the X and X authors, and took place from July to December 2023. All interviews took 
place virtually, using video conferencing software, and were transcribed using a speech-to-text transcription service (with 
additional transcription and amendments by the lead interviewer). One interviewee did not consent to their interview quotes 
being used in this paper.

\subsection{Data analysis}
For  our  data  analysis,  we  adopted  a  grounded  theory  epistemology,  allowing  us  to  surface  latent  themes,  patterns,  and 
social interactions from our data \cite{charmaz2023constructing}. Four authors contributed to the coding process using an inductive approach, allowing 
us to derive codes based on concepts from the data. We used Atlas.ti data analysis software to support the process. We 
completed a round of asynchronous coding, generating an initial codebook of 102 codes. We then deliberated and identified 
areas of consensus/dissensus and overlap, enabling a synthesis of the codebook down to a total of 16 codes. Each interview 
was then coded again by two separate authors using this revised codebook. See the codes in Table 1 below:

\begin{small}
\begin{table}[h]
\caption{\label{summarydata}Codes}
\begin{tabular}{|c|c|}
\hline
Auditing service under LL144 & Auditor legitimacy \\ \hline
Background of auditor & Beneficiaries of the LL 144 regime \\ \hline
Comparison with other (audit) domains & Components of the audit \\ \hline
Data practices/issues/metrics & Enforcement of LL144 \\ \hline
Evaluating audit/auditor practice & History/context of the law \\ \hline
Incentives to comply with the law & Lessons for policymakers \\ \hline
Opinion on compliance & Opinions of the law
\\
\hline
\end{tabular}
\end{table}
\end{small}

\subsection{Limitation of study}
This paper reports on empirical research  on the implementation of LL 144, and its relationship to United States federal 
employment  law  and  civil  rights  law.  This  gives  our  research  a  narrow  geographical  focus.  We  believe  that  there  are 
generalisable lessons emerging from our findings that will be of utility to other jurisdictions/domains, but we also recognise 
that some of the findings (e.g. improvements in the law) are more applicable to the United States context (other nations do 
not, for example, use the four-fifths rule to measure disparate impact). A limitation for our interview findings is that not 
every  auditor  with  attributable  published  audits  under  LL  144  accepted  the  offer  of  participation.  Additionally,  the 
management consulting firms currently offering AI audits for compliance with other legislation targeting AI/data-driven 
systems (e.g. the European Union’s Digital Services Act) are (at the time of research) not conducting audits under the LL 
144  regime  and  are  therefore  not  represented  in  this  study.  Participants  were  not  offered  compensation  for  their participation  in  this  study.  See  Table 2  in  the  Appendices  for  Participant  ID,  including  background  and  organisational information, and 4.1 for the kinds of roles and services auditors enact in the LL 144 regime.

\section{BACKGROUND OF LOCAL LAW 144}
An introduction to the specifics of LL 144 is critical to engaging with our thesis. In this section, we present core definitions, 
the  central  tenets of  the  law,  and  insight  into  the  history  and  context  in  which the  law  emerges.    Significantly,  LL 144 
offers particular definitions of key terms; the same terms may be defined differently in other contexts.

\subsection{History of the law}
\begin{figure*}[!th]
  \centering
  \includegraphics[width=\linewidth]{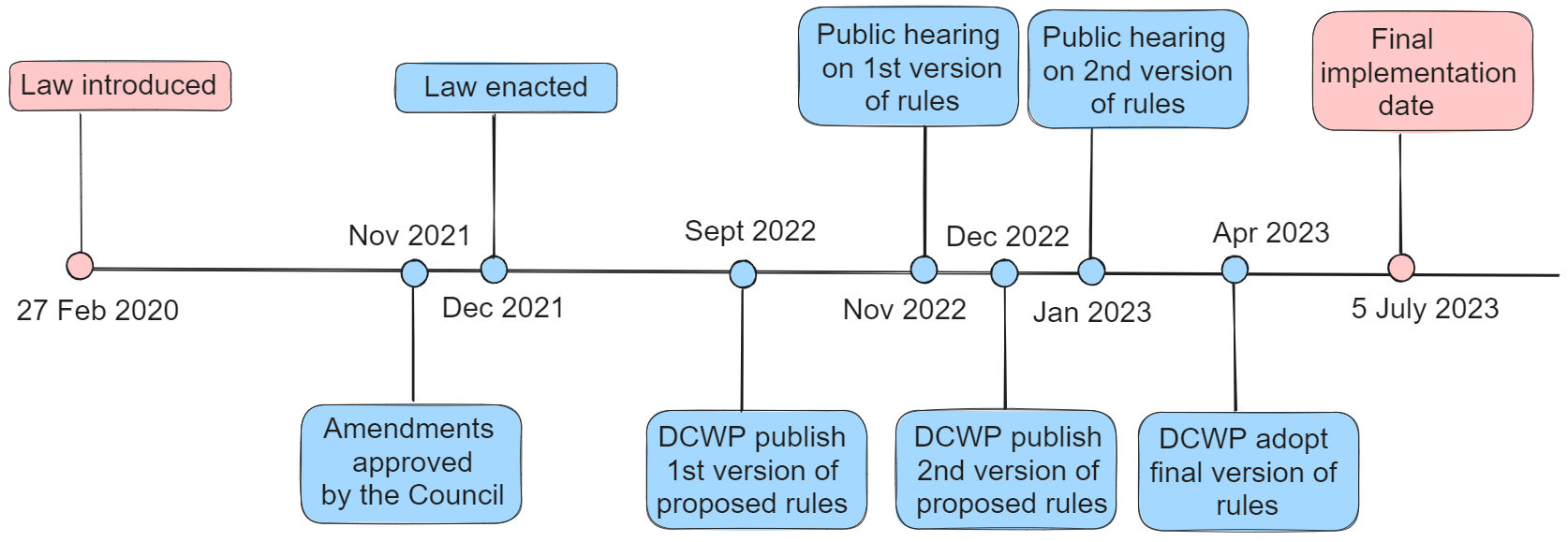}
   \caption{History of LL 144}
  \label{fig:fig1}
\end{figure*}

LL 144  is the result of a long-running effort by the municipal government of NYC, pushed by many civil rights advocates, 
to create a regulatory structure for algorithmic systems \cite{cahn_2021, Kirchner2023,Lohr2023}. This effort first focused on city government systems and 
eventually regulating only commercial algorithmic hiring systems. It was first passed by the NYC Council in late 2021 
\cite{NYCC2021} which tasked the Department of Consumer and Worker Protection (DCWP) with rule-making and enforcement. There 
were two revisions, with a period of open public comment, and multiple delays to the lawmaking process. The third and 
final version was adopted in April 2023 and implemented in July 2023\cite{NYC_AEDT2023}. See Figure ~\ref{fig:fig1} ‘History of the law’ for more 
details.

\subsection{Key aspects of the law and its context in US anti-discrimination law}
While  LL  144  marks  the  first  attempt  to  require  bias  auditing  of  AI  systems,  its  text  and  theory  of  change  awkwardly 
interacts with existing US employment anti-discrimination law. Some of the key aspects of this law are borrowed from 
existing litigation and standards—for example, definitions for race and gender and established formulas for assessing bias 
through impact ratios, as we detail below.  Some, such as the definition of AEDT, relied on law drafts, public comments 
and revisions to shape. We found that auditors were often involved in negotiating how exactly the novel definitions offered 
by LL 144 should be interpreted in light of established norms and laws. 
LL 144 requires only employers/hiring agencies that use the system to conduct a bias audit, but not the 
developer/vendor/platform  that  builds  or  sells  the  system \cite{NYC_DCWP2023}\footnote{The DCWP’s FAQ states that ``Employers and employment agencies are responsible for ensuring they do not use an AEDT unless a bias audit was done. The vendor that created the AEDT is not responsible for a bias audit of the tool” \cite{NYC_DCWP2023}.The decision to only impose obligations on end-users was primarily due to the city’s jurisdictional limits—developers of AEDTs or hiring platforms are unlikely to be domiciled in New York City. Additionally, employment law generally focuses on outcomes of decision-making, such as moments of hiring and firing, which is typically undertaken by employers and not by the platforms.}.  The  law  imposes two primary  obligations on  any  NYC employer using AEDTs for decisions on hiring and/or promotion.

\begin{itemize}
    \item \textbf{Annual `bias audit’ of the AEDT conducted by an independent auditor} for race and gender features and the 
intersection thereof. The bias audit report must then be posted on the employer’s website. The prescribed audit is more accurately described as a ‘disparate impact’ audit\cite{ajunwa2016hiring,barocas2016big,feldman2015certifying}, as it measures a specific form of algorithmic bias \cite{NYC_DCWP2023}. The law defines independent auditors as third-party experts who have no financial stake in the success of the 
product or the financial outcome of the employer.
\item \textbf{Transparency  notice  about  use  of  AEDT  and  right  to  opt-out  for  candidates.}  The  auditor’s  report  (see Figure ~\ref{fig:fig3}: Example audit report in the Appendix) must be posted in a publicly accessible location on their website, typically 
on a human resources or available jobs page; there is no mandate for audits to be submitted to the DCWP or a 
public  central  repository.  For  job  applicants,  this  disclosure  is  typically  appended  to  the  job  listing  on  the 
employer's website and/or recruitment platform(s) but can be delivered directly to the candidate.
\end{itemize}

LL 144 defines \textit{race and gender} using categories established by the EEOC \cite{EEOCrace}. Other protected categories central to 
AI fairness scholarship and employment civil rights law are not included in the audit requirement, such as disability, age 
and religious status. The law requires the audit report to present an impact ratio table. Figure 3 in the Appendix provides 
an  example  of  these  impact  ratio  tables.  The  DCWP  is  tasked  with  enforcing  these  rules  and  assessing  civil  penalties 
between \$500 and \$1,500 per day for violations\footnote{One auditor noted the enforcement penalties have some ambiguity about whether this fine applies per-candidate, which would drastically increase the cost,or per audit [P10].}. The jurisdiction of the law covers any position with a primary workplace 
in the boroughs of NYC, including remote work jobs that are headquartered there.

The law defines AEDTs so that many (perhaps most) of systems one might identify as `employment decision tools’ 
available on the market and offered by recruitment platforms are out of scope; it also grants employers significant discretion 
to  decide  if  their  systems  are  in  scope.  \textit{AEDTs}  are  defined  as  machine-learning/AI systems that `substantially assist or 
replace discretionary decision making’ in the hiring or promotion process \cite{koshiyama2021towards,leslie2019understanding}. \textit{Substantially assist} is defined as being 
the  primary/majority  reason,  or  predominant  reason  among  several,  for  a  `selection’  decision.  To  the  authors’  best 
knowledge, there is no system on the market that fully replaces human decision-making in hiring, and therefore the scope 
of the law is \textit{de facto} determined by the meaning of `substantially assist.’ However, `assisting’ is a feature of the human 
and organisational aspects of the hiring process, and not of the technical system. Because of this, two employers could use 
the same model from the same vendor, but have different interpretations of whether the output meets LL 144’s definition 
of ‘substantial.’ \textit{Machine learning/artificial intelligence} is defined in terms of outputting a simplified score or ranking on 
the basis of multiple feature inputs, in which the machine in part algorithmically weights those features. This narrows the 
scope further to deep-learning methods, and excludes even intentionally discriminatory weights that are set manually.  

The  law  draws  on  concepts  from  US  employment  law  to  determine  how  an  auditor  should  assess  for  bias.  The  law 
requires the creation of an impact ratio table, where an \textit{impact ratio} is a method for measuring discriminatory outcomes as 
the relative selection rate between demographic groups. A \textit{selection rate} is the frequency at which members of a group are 
chosen to move forward in a hiring/promotion process or rejected/screened out; `selection’ does not refer only to the final 
hiring decision, but also all decisions before that. LL 144 also makes use of a ‘scoring rate’ to measure impact ratios, 
capturing the fact that AEDTs are often used for producing simplified scores or rankings such as personality or intelligence 
scores. The \textit{scoring rate} considers the frequency at which the mechanism gives members of a group higher than median 
scores. Therefore, an impact ratio is a comparison of rates (not of absolute numbers), with the rate for one group as the numerator and the rate for the most-selected group as the denominator (the denominator can also be selection rate of the
entire population).\footnote{In practice, `less-selected’ group typically corresponds to a `historically-disadvantaged’ or `protected’ category. However, those are not synonymous. The
less-selected group could be a historically-advantaged/non-protected group, e.g., if an employer only hires members of a historically-disadvantaged group for low-status jobs. Thus it is important to recognize that disparate impact cannot measure historical discrimination.} An impact ratio of 1.0 means a perfectly equal selection rate between groups; an impact ratio less than one indicates a discriminatory outcome against the less-selected group, and the lower the fraction, the more discriminatory the outcome is.

The authors of LL 144 diverged from existing norms in US employment law when determining what action a company
must take upon discovering algorithmic bias. \textit{Disparate impact} (aka adverse impact) is a state of impermissible
discriminatory outcomes in US employment law. Disparate impact does not require intent to discriminate; rather, it captures
systemic discrimination for which the employer is nonetheless responsible, measured only by the outcomes of the
system\cite{selbst2019fairness}\footnote{That no intent need be inferred to measure discriminate outcomes is why disparate impact is often considered illustrative of algorithmic bias \cite{selbst2017disparate, feldman2015certifying}}.The four-fifths rule is a US anti-discrimination law convention, where disparate impact is defined as an impact
ratio below 0.8 (four-fifths), meaning the selection rate of the less selected-group is below 80\% of that of the most-selected
group (or entire population) \cite{fedReg_2024}. Rather than a rule or a law, the four-fifths rule operates as a guideline that plays the central
role in the EEOC’s decision tree for permissible selection procedures\cite{EEOCemployment}. An impact ratio that falls below 0.8 is subject
to regulatory scrutiny and must have additional justification to be legal. An impact ratio above 0.8 grants a presumption of
non-discriminatory outcomes, but is not an absolute shield from discrimination claims—other aspects of the employer’s
selection process may nonetheless be illegal. As critics point out, the four-fifths rule is a bureaucratic compromise that is
fundamentally arbitrary and not grounded in empirical evidence, and so may not be useful for judging algorithmic bias \cite{EEOCadverse,watkins2022four}.

Most crucially for auditors, LL 144 makes use of conventional measures of bias, yet sets no minimum for permissible
impact ratio rates. The law imposes no obligations to cease using any system that an audit discovers to cause disparate
impact\cite{NYC_DCWP2023}. All that is required is that the employer make the annual audit publicly available. The primary role the four-
fifths rule plays in LL 144 is its absence. Part of LL 144’s theory of change is the idea that a market for independent
algorithm auditing and transparency reporting will incentivise gradual adoption of better practices, thereby protecting
jobseekers’ civil rights. This theory of change espouses (and relies on) transparency, jobseeker autonomy, and reputational
pressure, as opposed to placing limits on companies or vendors using AEDTs or proactive investigations by regulators (P4,
P5). While an algorithm auditor might conceive of their role as helping clients reduce discriminatory outcomes and protect
civil rights—via services such as consulting, model design advice, governance practices, legal guidance, etc.—their
mandated role is only the production of audit reports. Absent a mandate to remediate discriminatory outcomes from LL
144, auditors must find secondary avenues to achieve their goals and provide valuable services to clients.

\subsection{Analysis of the public comments}
As part of our initial background research, we conducted a qualitative analysis of the third and final round of public
comments on the DCWP’s rule-making process, submitted in January 2023\cite{deis1992determinants}. Our analysis revealed disagreements
primarily about the scope of the law (which systems would be subject to it) and the positionality of auditors (does the
legitimacy of audits require the independence of auditors?). Of the 46 total comments, we identified 13 civil society
organisations (including some authors of this paper), four audit service firms, six employment law firms, seven system
developers, and three employment platforms\footnote{Employment platforms help jobseekers, recruiters, and employers rank and source jobs and/or candidates. They are out of scope of LL 144 as they do not
make final hiring decisions.}. Labour rights advocacy organisations, including labour unions, were not well represented, with the bulk of the civil society comments focusing on issues of algorithmic accountability. Only a few
comments specifically detailed the consequences for job seekers of the details in the law.

Generally, both hiring platforms and AEDT vendors expressed concern about the law being too broad and onerous,
with civil society stating the reverse, arguing industry actors may try to skirt regulation. We found evidence that AEDT
vendors lobbied the government for narrower definitions, language that allowed for exceptions, and higher degrees of
control over the audit process. One platform suggested removing the requirement for an independent auditor in favour of
an internal team, while a vendor argued the 10-day transparency notice requirement would burden NYC employers. Audit
service/audit tooling firms were in favour of independent auditing on grounds of the trustworthiness of audits and market
incentives. In official comments and reporting, civil society organisations were split on whether LL 144 should be revised
or scrapped. One local coalition opposed the law due to its limitation to hiring decisions, calling for regulatory scrutiny on other forms of algorithmic employment tools, like work assignments or discipline \cite{cahn_2021}. Civil society commenters frequently called for the removal of the ‘substantially assist’ criterion, arguing it would allow employers to declare themselves out of scope—effectively nullifying the law. In contrast, industry commenters stated that it should remain, which it did. The tensions between groups in the public comments highlights the role that relational dynamics between industry, civil society, and the DCWP played in shaping the final law.

\subsection{Public availability of audit reports}
Despite the law’s emphasis on transparency, very few audits are available publicly in the first six months of final rule
implementation. We searched for available audits using ad hoc search methods prior to conducting interviews, we identified
19 audits claiming to be related to LL 144, including five from system developers/vendors not subject to audit under the
law. Some of these audits are no longer available online. Both follow-up systematic searches four months after
implementation in a companion project [Redacted, forthcoming] and an open-source collection assembled by researchers
at the New York branch of the American Civil Liberties Union \cite{Gerchick2023} found commensurate results. Surveys indicates that
AEDTs of all varieties are thought to be widespread, but there is no available data about the baseline rate of adoption of
AEDTs generally, let alone adoption of systems that fall within the scope of LL 144 \cite{Indeed2023,Pew2023,selbst2019fairness}. Therefore, it is not possible to know precisely how many audits we should expect to find, and lacking a central repository, it is also impossible to know precisely how many audits are published. However, given the many thousands of employers in NYC and widespread use of AEDTs, one would reasonably anticipate that more than a handful of employers would fall under scope of this law.

Under LL 144, employers have almost-total discretion to make a judgement on whether their systems are in scope for
multiple reasons, and jobseekers have no formal mechanism to challenge that decision. The absence of an audit on an
employer’s website neither implies the absence of an AEDT (as they may use an AEDT self-determined as out of scope),
nor non-compliance (they may use a genuinely out of scope AEDT or none at all), nor even a lack of audit (the process
may have been conducted, but the report is not published). Though the audit reports themselves were not a focal point for
our study, mapping the audits and observing these conditions was informative for our interviews with auditors.

\section{FINDINGS FROM INTERVIEWS}
Our initial investigation into the public comments that shaped the law revealed relational tensions and potential gaps in the definitions created by the law. While public comments were split on whether the law was too narrow, or too broad, we
wanted to gather information from auditors about the actual implementation of the law. Informed by our investigation into
the law, we asked our interview participants questions roughly following four themes:

\begin{itemize}
    \item \textbf{High-level organisational questions}, including questions on the auditor team and their experience
    \item \textbf{Relational dynamics} between auditor, companies audited, AEDT tool developers and regulators, including
questions on how key information is communicated
    \item \textbf{Audit methodologies}, including questions on the content, process, artefact of audit and success criteria
    \item \textbf{Learnings from the audit process or broader LL 144 regime} for use by policymakers
\end{itemize}

\subsection{Opinions of the law overall were that it was well-intentioned but seriously flawed}

\textit{`I don’t think [LL 144] should be admonished for not being perfect. I call this the first pancake: without the first pancake, none of the other pancakes would be better. And the first one is always pretty awful in the pan, no
matter what you do.’ [P6]}\\

We asked our interviewees for their opinions of the law and the law-making process, which revealed a consensus about
some fundamental challenges that impeded auditing practices. While most auditors we interviewed had a positive opinion
about the spirit of the law, they also agreed that it was not a particularly effective auditing mechanism. Participants gave strong opinions on employers’ discretion to decide if systems meet the `substantially assist’ definition (see 2.3): `[the AEDT definition] creates huge loopholes’. Participants with knowledge of the law-drafting process pointed to lobbying efforts by
both AI vendors and large employers to water down the definition. Many auditors [P2, P5, P14] noted a key challenge with
the legislation is that ``it's too open for interpretation when it comes to the scope of the systems” [P2]. As [P5] put it, the ``first hurdle” many auditors experienced with this law was their clients asking if it was a good use of resources. Some auditors did not like the law’s theory of change. The expressed a view that it does not do enough to protect job applicants because audits and notices to applicants, while important, will not enable them to make an informed decision about whether to be subject to an AEDT or a human reviewer, or about whether to apply for the job at all [P11, P13]. Others [P1] noted that the audit reports can serve as the basis for civil lawsuits or regulatory action against a company or vendor if an audit shows the tool may be biased against a certain protected demographic, but P4 noted that this theory of change doesn’t account for the impossible burden of proof an injured job-seeker would face in making a successful complaint to the DCWP. In their view, the burden of proof for discrimination needs to be revised for this theory of change to work.

\subsection{Auditing regimes create a variety of auditing roles and services}
Our interviews reveal little shared agreement of the role and remit of an `auditor’ of AEDT systems in this regime, and as
an algorithm auditor more generally. Rather than identifying one single function or role of an auditor, we found participants broadly self-identified into one of four categories of audit ‘actors’ that offered different services (or a mixture of these services) to help clients comply with LL 144:

\begin{itemize}
    \item Participants that offered a `pre-audit’ service to help a company become ‘audit ready,’ including helping with their
    data collection process, providing data governance tools, or other frameworks [e.g. P1, P5, P6]
    \item Participants that conducted the audit and wrote the report for a client [P2, P3, P7, P8, P11, P12]
    \item Participants that offered additional guidance and mitigation strategies for the company [P9, P10, P13, P14, P15]
    \item Participants who offer a service to certify that an audit has been conducted in an appropriate manner [P11, P12]
\end{itemize}

Clients received different services when conducting an audit, depending on the auditor. For example, participants from law
firms [P3, P16] might offer a judgement about whether the client is compliant with the law and offer strategies to become
compliant. Additionally, while the law plainly imposes obligations on only employers, the DCWP does offer secondary
guidance that vendors ‘may’ conduct an independent audit for their own purposes (presumably for market advantage)—a phenomenon observable in the various collections of audit reports—which means that auditors must also consider the needs
of clients with different regulatory statuses \cite{NYC_DCWP2023}. These differences in services can also be observed in the public audits
artefacts, which vary widely between the minimal quantitative measures of impact ratios required by the law, to far more
extensive assessments of algorithmic accountability processes and audit readiness\footnote{For examples, see the earlier cited open-source repository, Gerchick \& Watson (2023).}. The lack of clear roles, responsibilities, and services around the law furthers the need for additional relational work to comply with the law. Interpretation and implementation are left to auditors and their clients, resulting in fractured roles and responsibilities.

\subsection{Disagreement over what constitutes a legitimate auditor and legitimate auditing practice}

\textit{`I think it’s important that we are not auditors, because auditors by our interpretation of that definition, are an
independent authority … nothing that an auditor does should be taken as advice. We see them as an assessor that can
validate, because that’s the primary function of an auditor.’ [P6]}\\

The algorithmic auditing ecosystem is nascent and currently lacks the structure and standards of other audit domains, such
as financial services \cite{raji2022outsider}. Auditors who provide services to LL 144 clients are both interpreting the context of that law,
and conceptualising what a sustainable algorithmic auditing business may look like in the long-run—how might they
provide multiple forms of utility to their clients and persuade potential future clients of their value? This relational dynamic
often strains the traditional meanings of `audit.’ Participants expressed different views on what practices or qualities made
them legitimate auditors under LL 144. One interviewee cited that the diverse input involved in their ``crowdsourced,
participatory” [P1] audit certification service would raise legitimacy questions, while another stated their “transparent
approach” [P12] defined them as a legitimate auditor practice. One interviewee said that the law is relatively clear on the
components of the bias audit itself, but not on exactly \textit{who} is the optimal actor to conduct it.

Interviewees cited their domain expertise as one factor that qualified them to conduct audits. These included social
statistics [P7], software engineering [P2], experience with HR [P4] or occupational psychology [P10]. In lieu of formally
recognised standards for algorithm audit, our participants described drawing on internal and external codes of practice. For example, one organisation [P9] described aligning themselves with existing assurance standards, such as the ISAE 3000
issued by the International Auditing and Assurance Standards Board \cite{ISAE2013}, as well as with draft audit criteria supplied by
another organisation that participated in this study. Many of our interviewees reported conducting audits for companies
they had pre-existing relationships with. In other auditing domains, such as financial audit, rules around the auditor-client
relationship are tightly controlled \cite{SHRM2022}. Two participants felt that the financial incentive in providing a bias audit under LL
144 might create challenges for independence [P1] [P12]. Several participants suggested that more flexibility about
independence would increase the effectiveness of the bias audit: `I needed [to work closely with] our clients to have better
data governance practices in order for me to be able to do a better audit for them’ [P14].

Others reported scepticism about the extent to which codes of practice for auditors under LL 144 appear to divert from
existing practice elsewhere. For example, two participants [P10, P1] highlighted that the US Sarbanes Oxley Act 2002
strictly prohibits independent auditors from providing `certain non-audit services’ to clients \cite{US_sarbanes}. Our interviews also
revealed that some early LL 144 audits were being conducted in-house by employers on the grounds that another
department within their organisation would qualify as independent. This touches on a tension between several interview
subjects: some felt offering additional advisory services was not in opposition to the law’s definition of being an
independent auditor [P3]. But others, like P5 and P6, felt a core component of being an independent auditor was that they
could not provide advice or guidance on how to mitigate these issues. In their view, an auditor should be a one-off assessor of an AI system to provide ``a review, a score, or a grade” that validates or certifies a company’s compliance with LL 144 [P6]. They also distinguished the role of an auditor as specifically assessing an AI product at a particular point in its lifecycle, as opposed to a continuous form of evaluation. This is why some auditors, like P5 and P6, explicitly rejected the term `auditor’ to describe their own services. From a business perspective, P5 and P6 also felt that offering companies the service of becoming ‘audit ready’ provided more opportunities to help their clients comply with other kinds of laws like the EU AI Act or the EU’s Digital Services Act. Like other elements of the law, what exactly constitutes a legitimate auditor remains vague and undefined within the law itself, other than the minimal requirement of financial independence from the outcome of the audit report. Therefore, the responsibility to define auditor roles falls to employers and auditors themselves. Without clarity on the role of auditors, in practice, the power dynamics between existing actors in the field establish a working practice for implementation.

\begin{figure*}[!th]
  \centering
  \includegraphics[width=\linewidth]{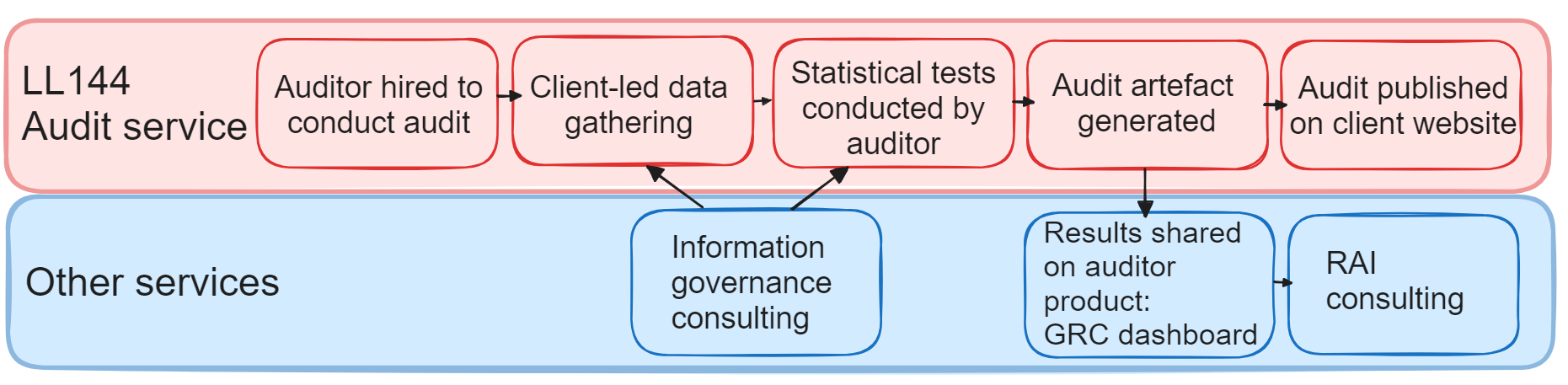}
   \caption{Services offered under the LL44 regime by auditors}
  \label{fig:fig2}
\end{figure*}

\subsection{LL 144 created an opportunity for more companies to adopt wider responsible AI practices}

\textit{`This is your golden opportunity to change things in your organization. That’s how we try to put it.’ [P6]}\\

Several of the participants we interviewed offered their clients additional services beyond those required under LL 144.
P5, P6, P9, and P12, for example, all noted they offer wider responsible AI consultancy services, including support with
data governance, wider AI ethics training, and even a SaaS platform to help clients enact responsible AI practices. Others,
like P16, noted that they offer legal advice along with the audit. Others, like P1, P5, P6, P7, and P8, noted they are advising clients on how to get compliance-ready for other possibly forthcoming laws. Several participants noted that LL 144 created an incentive for companies and vendors to adopt wider responsible AI and data ethics practices, enabling participants to offer additional responsible AI services. P5 and P6 noted this kind of work was more lucrative than just helping companies conduct an audit. P14 noted that the law, while highly imperfect, had the effect of getting their clients to think about responsible AI and embed new language and ways of thinking into their corporate culture. According to P14, the audits ensured that companies had a single AI system document and that their role as an auditor was to provide documentation and a `grounded language’ around responsible AI. The creation of a document enabled the wider staff to understand ``that it's absolutely appropriate for them to have a say in how AI is developed and applied in their company” [P5].

\subsection{Auditors faced numerous obstacles around data practices, metrics and cultural attitudes toward data with
vendors/clients}
While technically straightforward, the audit process presents auditors with challenges concerning data practices and
cultural attitudes in their engagements with vendors and clients. The question of how to procure the necessary data has
become a source of complexity for auditors. LL 144 only imposes obligations on the end-user, the employer, creating a
relational dynamic between employer and vendor (with auditors mediating in the middle). This requires the end-user and
the auditor to persuade the vendor for data access per the legal obligations\footnote{The law is ambiguous about the extent to which an auditor may use a vendor’s own audit. `Pooled’ data from the vendor is allowed, as long as an employer also contributes their own data to the pool. See Raghavan et al. (2019) \cite{raghavan2020mitigating} on tradeoffs using pooled data in hiring algorithms}. Even if the data is available, auditors encounter
significant hurdles in accessing it due to vendor concerns around releasing proprietary systems or sensitive data. P14 noted
that vendor cooperation was ``key” and ``one of the biggest obstacles to clients complying with the NYC law.” According
to P14, ``our only adversarial or third party was the vendor. So we had to kind of take a very hard stance, getting our
position and be kind of aggressive with the vendor to get us the data that we wanted so that we could produce that audit
the way it's meant to be.” P8 and P7 also referred to the role of non-disclosure agreements between clients and vendors as
an obstacle that prevented access to the data necessary to do the audit. Often, auditors use tools like APIs or databases that
allow controlled access to limited, but relevant, data. This tension between auditor and vendor not only extends the duration
of audits, but also underscores the importance of establishing streamlined processes for data sharing and accessibility.

Many participants cited a lack of guidance from DCWP here as a major contributing factor to data issues. As P14
noted, `the guidance still has a lot of discretion and flexibility,” and could be interpreted by different firms and vendors in
different ways. This included guidance on how to select data for the audit; according to P14, ``[the guidance] almost doesn't
talk about [sampling data].. So I mean, if we wanted to be in bad faith, we could look at our clients dataset, and we could
very easily pick rows that would give stellar results.” While P14 and other auditors acknowledged they would never do
this, the ease of cherry-picking or biasing audit results, raising a serious risk to the legitimacy of the law. It also reduces
the reliability of any measure of the success of the law that relies on tracking changes to impact ratios.

Many clients and vendors were initially resistant to acknowledging biases within their systems, making it difficult for
auditors to help them understand their potential biases and potential mitigation strategies. P2 noted that some clients were
``frugal” with their data but would provide full access after initial results indicated an issue. P14 described how the process
of conducting the audit was similar to ``the stages of grief.” They reported stages of denial (``oh, no, we must have given
you the wrong data or something like that”), despair (``oh, no, like, this is the worst thing in the world that could happen to
us”), and finally acceptance from the client. This sense of denial matched the perspective of P8, who described clients who
didn’t collect data on ethnicity or gender ``because it's sensitive,” and then incorrectly claimed that this lack of data
collection meant that bias was impossible.\footnote{Federal regulations do require employers to keep anonymized demographic records of their hiring practices, recorded on the EE-01 form.} Bridging this gap between denial and acknowledgment proves to be a delicate
and recurring challenge in the audit journey. The issue of `imputed or inferred' demographic data adds another layer of
complexity. While explicitly prohibited under LL 144 regulations, auditors noted that this practice persists, resulting in
auditors often using names to infer demographic information like race, ethnicity, and gender. One auditor described a case
where a client lacked any data on gender but made it up by assuming they had a 50/50 split because the national population
in the USA is about a 50/50 male/female split. These instances of imputation/inferral are particularly problematic in diverse urban settings like NYC and raise questions about the validity of audits conducted under LL 144.

Additionally, the statistical test that LL 144 required auditors to use to determine disparate impact raised obstacles. P14
noted that this test required using `the most favoured group as the reference group, and any group that scores less favourably, those are your protected groups,’ conflicted with `historical’ ways of doing disparate impact analyses where
the protected group is pre-selected, before viewing the data. As a result, this caused the statistical validity of the test to go `out the window’ and enabled `cherry picking’ of data to tell a story that the system was not biased. Several auditors noted that having a mandatory, standardised metric for protected groups is important, but noted that a mandatory metric must be purpose-specific and appropriate [P6]. P10 noted the metric offered by DCWP changed part way through the implementation of the law from the average to the median, which caused delays and additional work.

\section{DISCUSSION}
Our interview findings and analysis of the law lead us to conclude that despite some modest successes in socialising bias
audits among private companies, the law has fallen far short of its ambition to generate an effective bias audit regime that protects job applicants from discriminatory algorithmic tools. Of the thousands of employers in NYC, we could find only 19 who have publicly posted audits. While not a primary finding from this paper, we argue that the context of extremely low compliance is important for policymakers to note, both for making a judgement on the effectiveness of this auditing regime and for the wider implications for AI regulatory initiatives that may be proposed in the future. Below, we discuss four recommendations that policymakers seeking to develop similar algorithm auditing regimes should consider.

\subsection{Auditing laws must have a theory of change that holds developers accountable, makes audits the source of
action/enforceability and prioritises jobseekers}
Councilwoman Laurie Cumbo has stated her motivations for sponsoring this law were to curb unjust hiring practices and
promote equality\cite{Ivanova2020}. In practice, however, the lack of audits and notices (and the poor readability/interpretability of those
that do exist) means job candidates may not directly benefit. Our interviews do not reveal evidence of candidate outcomes
being an influential driver or consideration behind the audits—according to P5, ``the needs of the [candidate] are so different
from the needs of an enterprise…and although we care about the [candidate], I don't think that it is incumbent upon [us] to
be able to satisfy the needs of that [candidate] at the same time as satisfying the needs of the enterprise.”

At present, the LL 144 auditing regime is insufficient for creating accountability relationships between jobseekers and
deployers or vendors: the lack of prescribed mitigation steps in the event of adverse impact means a candidate would need
to rely on plaintiff litigation for further action. Additionally, there is no requirement for the vendor to take action; an
employer would need to convince vendors to improve the fairness outcomes of a discriminatory system. Consequently, the
law forms a relational regime for industry actors to navigate – employers who use AEDTs must navigate between their
responsibility to LL 144, the potential censure of the EEOC, and the social cost of publishing a non-favourable bias audit.

To remedy this, we recommend future auditing laws take an ecosystem view to accountability, understanding that there
is `distributed responsibility’ \cite{cobbe2023understanding} between vendors, users/deployers and other actors. In practice, this might involve placing
audit requirements on vendors or hiring platforms (we note that this was complicated in the NYC context due to the city’s
jurisdictional limits). One of our participants proposed that vendors should be obliged to publish the audits of all employers
using their tools. For future regimes, it is critical that a regulatory body hosts a central audit repository, which will assist
candidates in making comparisons between different employers, law firms, and regulators in enforcement. Future laws
could mandate the disclosure of additional context about the employer and the tool (including contact details), which could
make the audit results easier for researchers and the public to parse and scrutinise. Audits could also be supplemented by
other kinds of responsible AI practices, like algorithmic impact assessments, that seek to address accountability issues in
different parts of the product development lifecycle \cite{Groves2022,selbst2017disparate}. Lastly, to avoid the over-representation of industry voices in public consultations around auditing regimes, it is critical for policymakers to intentionally engage with stakeholders
who will be impacted by these systems.

\subsection{Auditing laws need clear and meaningful metrics and definitions of key terms}
According to our participants, a major source of NYC LL 144’s failure is the last-minute addition of language around
AEDTs that `substantially assist’, which drastically reduced the number of companies in scope. Additionally, the statistical
metric used by the NYC LL 144 law (disparate impact, based on the four-fifths rule) is heavily rooted in US law and is not
applicable to other jurisdictions outside of the USA. Policymakers creating similar regimes must ensure they carefully
consider the terms and metrics used and ensure that these align with their local cultural context. Terms should always be
defined in a way that is weighted in favour of protections for those impacted by these technologies over those building
them. Regulators or departments responsible for overseeing these regimes can take steps to issue updated guidance about
the use of these tools. Crucially, policymakers must also not shy away from enforcement, which is a primary driver of
compliance in these regimes.

\subsection{Establishing an auditing ecosystem requires more standards for and oversight of auditors}
Many of the observed challenges surrounding the regime are about limited evaluation procedures and oversight and
enforcement capacity. Auditors are largely left to their own devices in defining ‘good’ practice in this space, and
interviewees suggested that the DCWP may be ill-equipped to enforce the law. Further, without a pool of certified or
accredited auditors, under this regime, any organisation without a financial stake in the AI system or audited company can
conduct audits. Even in good faith, this leaves the door open for inconsistencies in practice and prevents oversight that
might enable ecosystem-level learning.

Our interviews have also highlighted that auditing regimes for AI systems can lead to a range of actors delivering
various services beyond the strict definition of an `independent auditor.’ These include companies that provide the service
of making companies ‘audit ready’ or advising them on steps they can take to improve their data and responsible AI
practices. Our interviews suggest these roles are essential, but they must be clearly defined and differentiated in an auditing
regime. Future auditing regimes need to establish clear requirements and standards for what constitutes an ‘independent’
auditor. A next step for national governments would be to define standards of practice for algorithm auditors, for the benefit
of auditors and regulators. These standards should also set out methodological details surrounding the approach to auditing
for second- and third-party audits. This could build on, for example, the UK Information Commissioner’s Office guidance
for AI auditing \cite{ICO2020}. As suggested elsewhere \cite{raji2022outsider,koshiyama2021towards}, the financial auditing model could also provide a steer: for example,
an algorithm auditing equivalent to the US Public Company Accounting Oversight Board \cite{PCAOB} could be established. We
also draw attention to emerging non-governmental bodies with a remit to define algorithm audit standards, such as the
newly established International Association of Algorithm Auditors \cite{IAAA}. Previous work on AI accountability and assurance
has identified a need for an `AI ombudsman’ \cite{Jones2023} as an authority to field complaints and grievances. We propose that a
similar model could be adopted to mediate AI auditors. Additionally, increased regulatory capacity at the local and national levels is required for regulators to enforce audit compliance.

\subsection{Auditing laws must enable smoother data collection for performing tests and providing access to data for
certified auditors}
Our findings underscore the need for legal frameworks that facilitate seamless data collection, empowering auditors to
conduct meaningful assessments. One prevailing approach relies on auditors fostering positive relationships with a
14
system’s developers, emphasising concerns that evaluations often hinge on selective disclosure, potentially favouring
auditors who portray a positive image of the system. The issue of access is further complicated by a lack of trust in
companies releasing only favourable data, a challenge exacerbated by the fact that APIs are often developed and managed
by large, for-profit entities. In recent years, companies providing tools like APIs, intended for public system data use, have
raised costs and restricted tool usage. Notably, platforms such as Reddit have introduced charges for API access, while
others have shut down their APIs altogether. Consequently, auditors frequently find themselves obligated to declare their
intent for data usage, introducing uncertainty and potential bias into the auditing process. From a legal perspective, auditors
face legal vulnerability due to existing language within privacy and anti-hacking laws. For instance, the US Computer
Fraud and Abuse Act criminalises ‘unauthorised access to systems, particularly regarding data collection without
permission \cite{DOJcfa}. The ambiguity surrounding methods classified as authorised versus unauthorised creates uncertainty when
auditors use tools like data scrapers to access system data, potentially violating platform Terms of Service. In response to
these challenges, we advocate for legal clarity and protection for algorithm auditors to conduct thorough and meaningful
evaluations. Precision in legal language, including terms of use for data collection methods and tools, is essential to
ensuring that auditors are shielded from unwarranted legal repercussions.

\section{CONCLUSION}
In this paper, we explore the implementation LL 144, the first law creating an algorithm bias audit regime. While an
admirable first attempt, its vague definitions and narrow scope ultimately failed to drive a robust implementation of bias
audits, leaving the real work of auditing to be decided by industry actors, auditors, and hiring platforms. The law created
a warped incentive structure resulting in under-compliance for employers whose systems should be in scope, but also over-
compliance from vendors and platforms that are not in scope. This is in part a consequence of placing all formal
responsibility on the software’s end user—the employer using the AEDT—even though the end-user typically does not
build or even have access to the aspects of the system that one would typically require for a robust algorithmic audit. In
addition, the law grants the end-user total discretion over whether their system is in scope, and offers multiple loopholes
for employers to move out of scope without addressing discriminatory outcomes. While the law specifies what the audit
should include, it fails to specify the role of the auditor, creating confusion about the appropriate credentials and practices
of independent auditors. Consequently, there is little standardisation in LL 144 audit services. The law has driven a
secondary market of responsible AI services that, in many cases, are becoming the primary activity under this regime.
Auditors are challenged by navigating the accountability relationships created by the law, including negotiating system
access from developers and with legal counsel about whether a system is in scope. Clarifying definitions and the
responsibilities between vendors, employers, and jobseekers may help establish clearer accountability relationships and
mitigate potential non-compliance. We join calls from across the FAccT community that similar laws should address the
capacity building of algorithm audit expertise, standards of practice, and robust oversight mechanisms \cite{raji2022outsider}. For mandated
audits to become a meaningful accountability mechanism in AI, it is critical that these regimes apportion accountability in
a way that delivers beneficial outcomes for people impacted by these systems.

\begin{acks}
This project was supported in part by the Open Society Foundation and Siegel Family Endowment. The paper benefited significantly from a related project conducted in collaboration with researchers from the Cornell University Citizens and Technology Lab, lead by J. Nathan Matias. The authors would like to thank Catherine Gregory, Sohaib Malik, Rebecca Ghani, Octavia Reeve, and Ranjit Singh for their thoughtful comments and revisions to this paper.
\end{acks}

\bibliography{bibliography}


\begin{thebibliography}{49}


\ifx \showCODEN    \undefined \def \showCODEN     #1{\unskip}     \fi
\ifx \showDOI      \undefined \def \showDOI       #1{#1}\fi
\ifx \showISBNx    \undefined \def \showISBNx     #1{\unskip}     \fi
\ifx \showISBNxiii \undefined \def \showISBNxiii  #1{\unskip}     \fi
\ifx \showISSN     \undefined \def \showISSN      #1{\unskip}     \fi
\ifx \showLCCN     \undefined \def \showLCCN      #1{\unskip}     \fi
\ifx \shownote     \undefined \def \shownote      #1{#1}          \fi
\ifx \showarticletitle \undefined \def \showarticletitle #1{#1}   \fi
\ifx \showURL      \undefined \def \showURL       {\relax}        \fi
\providecommand\bibfield[2]{#2}
\providecommand\bibinfo[2]{#2}
\providecommand\natexlab[1]{#1}
\providecommand\showeprint[2][]{arXiv:#2}

\bibitem[IAA({[n.\,d.]})]%
        {IAAA}
 \bibinfo{year}{[n.\,d.]}\natexlab{}.
\newblock \showarticletitle{IAAA - International Algorithmic Auditors Association}.
\newblock  (\bibinfo{year}{[n.\,d.]}).
\newblock
\urldef\tempurl%
\url{https://iaaa-algorithmicauditors.org/}
\showURL{%
\tempurl}


\bibitem[fed(2024)]%
        {fedReg_2024}
 \bibinfo{year}{2024}\natexlab{}.
\newblock \showarticletitle{Code of Federal Regulations}.
\newblock \bibinfo{journal}{\emph{29 CFR Part 1607 - General Principles}} (\bibinfo{year}{2024}).
\newblock
\urldef\tempurl%
\url{https://www.wired.com/story/opinion-new-york-citys-surveillance-battle-offers-national-lessons/}
\showURL{%
\tempurl}


\bibitem[Ajunwa et~al\mbox{.}(2016)]%
        {ajunwa2016hiring}
\bibfield{author}{\bibinfo{person}{Ifeoma Ajunwa}, \bibinfo{person}{Sorelle Friedler}, \bibinfo{person}{Carlos~E Scheidegger}, {and} \bibinfo{person}{Suresh Venkatasubramanian}.} \bibinfo{year}{2016}\natexlab{}.
\newblock \showarticletitle{Hiring by algorithm: predicting and preventing disparate impact}.
\newblock \bibinfo{journal}{\emph{Available at SSRN}} (\bibinfo{year}{2016}).
\newblock


\bibitem[Barocas and Selbst(2016)]%
        {barocas2016big}
\bibfield{author}{\bibinfo{person}{Solon Barocas} {and} \bibinfo{person}{Andrew~D Selbst}.} \bibinfo{year}{2016}\natexlab{}.
\newblock \showarticletitle{Big data's disparate impact}.
\newblock \bibinfo{journal}{\emph{California law review}} (\bibinfo{year}{2016}), \bibinfo{pages}{671--732}.
\newblock


\bibitem[Bennett et~al\mbox{.}(2013)]%
        {bennett2013customer}
\bibfield{author}{\bibinfo{person}{Victor~Manuel Bennett}, \bibinfo{person}{Lamar Pierce}, \bibinfo{person}{Jason~A Snyder}, {and} \bibinfo{person}{Michael~W Toffel}.} \bibinfo{year}{2013}\natexlab{}.
\newblock \showarticletitle{Customer-driven misconduct: How competition corrupts business practices}.
\newblock \bibinfo{journal}{\emph{Management Science}} \bibinfo{volume}{59}, \bibinfo{number}{8} (\bibinfo{year}{2013}), \bibinfo{pages}{1725--1742}.
\newblock


\bibitem[Brennan et~al\mbox{.}({[n.\,d.]})]%
        {Brennan_AI_assurance}
\bibfield{author}{\bibinfo{person}{Jenny Brennan}, \bibinfo{person}{Lara Groves}, \bibinfo{person}{Elliot Jones}, {and} \bibinfo{person}{Andrew Strait}.} \bibinfo{year}{[n.\,d.]}\natexlab{}.
\newblock \bibinfo{title}{AI Assurance?}
\newblock
\newblock
\urldef\tempurl%
\url{https://www.adalovelaceinstitute.org/report/risks-ai-systems/}
\showURL{%
\tempurl}


\bibitem[Buolamwini and Gebru(2018)]%
        {buolamwini2018gender}
\bibfield{author}{\bibinfo{person}{Joy Buolamwini} {and} \bibinfo{person}{Timnit Gebru}.} \bibinfo{year}{2018}\natexlab{}.
\newblock \showarticletitle{Gender shades: Intersectional accuracy disparities in commercial gender classification}. In \bibinfo{booktitle}{\emph{Conference on fairness, accountability and transparency}}. PMLR, \bibinfo{pages}{77--91}.
\newblock


\bibitem[Cahn(2021)]%
        {cahn_2021}
\bibfield{author}{\bibinfo{person}{Albert~Fox Cahn}.} \bibinfo{year}{2021}\natexlab{}.
\newblock \showarticletitle{New York City’s Surveillance Battle Offers National Lessons}.
\newblock \bibinfo{journal}{\emph{Wired}} (\bibinfo{year}{2021}).
\newblock
\showISSN{1059-1028}
\urldef\tempurl%
\url{https://www.wired.com/story/opinion-new-york-citys-surveillance-battle-offers-national-lessons/}
\showURL{%
\tempurl}


\bibitem[Center(2023)]%
        {Pew2023}
\bibfield{author}{\bibinfo{person}{Pew~Research Center}.} \bibinfo{year}{2023}\natexlab{}.
\newblock \showarticletitle{AI in Hiring and Evaluation of Workers: What People Think}.
\newblock  (\bibinfo{year}{2023}).
\newblock
\urldef\tempurl%
\url{https://www.pewresearch.org/internet/2023/04/20/ai-in-hiring-and-evaluating-workers-what-americans-think/}
\showURL{%
\tempurl}


\bibitem[Charmaz(2023)]%
        {charmaz2023constructing}
\bibfield{author}{\bibinfo{person}{Kathy Charmaz}.} \bibinfo{year}{2023}\natexlab{}.
\newblock \bibinfo{booktitle}{\emph{Constructing grounded theory: A practical guide through qualitative analysis}}.
\newblock \bibinfo{publisher}{sage}.
\newblock


\bibitem[Cobbe et~al\mbox{.}(2023)]%
        {cobbe2023understanding}
\bibfield{author}{\bibinfo{person}{Jennifer Cobbe}, \bibinfo{person}{Michael Veale}, {and} \bibinfo{person}{Jatinder Singh}.} \bibinfo{year}{2023}\natexlab{}.
\newblock \showarticletitle{Understanding accountability in algorithmic supply chains}. In \bibinfo{booktitle}{\emph{Proceedings of the 2023 ACM Conference on Fairness, Accountability, and Transparency}}. \bibinfo{pages}{1186--1197}.
\newblock


\bibitem[Commission(2024)]%
        {EU_DSA}
\bibfield{author}{\bibinfo{person}{European Commission}.} \bibinfo{year}{2024}\natexlab{}.
\newblock \showarticletitle{The Digital Services Act package | Shaping Europe’s digital future}.
\newblock  (\bibinfo{year}{2024}).
\newblock
\urldef\tempurl%
\url{https://digital- strategy.ec.europa.eu/en/policies/digital-services-act-package}
\showURL{%
\tempurl}


\bibitem[Commission(2007)]%
        {EEOCemployment}
\bibfield{author}{\bibinfo{person}{Equal Employment~Opportunity Commission}.} \bibinfo{year}{2007}\natexlab{}.
\newblock \showarticletitle{Employment Tests and Selection Procedures}.
\newblock  (\bibinfo{year}{2007}).
\newblock
\urldef\tempurl%
\url{https://www.eeoc.gov/laws/guidance/employment-tests-and-selection-procedures}
\showURL{%
\tempurl}


\bibitem[Commission(2022)]%
        {EEOCrace}
\bibfield{author}{\bibinfo{person}{Equal Employment~Opportunity Commission}.} \bibinfo{year}{2022}\natexlab{}.
\newblock \showarticletitle{Definitions of Race and Ethnicity Categories}.
\newblock \bibinfo{journal}{\emph{Data Collection}} (\bibinfo{year}{2022}).
\newblock


\bibitem[Commission(2023)]%
        {EEOCadverse}
\bibfield{author}{\bibinfo{person}{Equal Employment~Opportunity Commission}.} \bibinfo{year}{2023}\natexlab{}.
\newblock \showarticletitle{Assessing Adverse Impact in Software, Algorithms, and Artificial Intelligence Used in Employment Selection Procedures Under Title VII of the Civil Rights Act of 1964}.
\newblock  (\bibinfo{year}{2023}).
\newblock
\urldef\tempurl%
\url{https://www.eeoc.gov/laws/guidance/select-issues-assessing-adverse-impact-software-algorithms-and-artificial}
\showURL{%
\tempurl}


\bibitem[Competition and Commission(2020)]%
        {ausCommission2020}
\bibfield{author}{\bibinfo{person}{Australian Competition} {and} \bibinfo{person}{Consumer Commission}.} \bibinfo{year}{2020}\natexlab{}.
\newblock \showarticletitle{Trivago misled consumers about hotel room rates}.
\newblock  (\bibinfo{year}{2020}).
\newblock
\urldef\tempurl%
\url{https://www.accc.gov.au/media-release/trivago-misled-consumers-about-hotel-room-rates}
\showURL{%
\tempurl}


\bibitem[Congress(2002)]%
        {US_sarbanes}
\bibfield{author}{\bibinfo{person}{US Congress}.} \bibinfo{year}{2002}\natexlab{}.
\newblock \showarticletitle{The Sarbanes Oxley Act}.
\newblock  (\bibinfo{year}{2002}).
\newblock


\bibitem[Council(2021)]%
        {NYCC2021}
\bibfield{author}{\bibinfo{person}{New York~City Council}.} \bibinfo{year}{2021}\natexlab{}.
\newblock \showarticletitle{A Local Law to amend the administrative code of the city of New York, in relation to automated employment decision tools}.
\newblock  (\bibinfo{year}{2021}).
\newblock
\urldef\tempurl%
\url{https://legistar.council.nyc.gov/LegislationDetail.aspx?ID=4344524&GUID=B051915D- A9AC-451E-81F8-6596032FA3F9&Options=Advanced&Search=}
\showURL{%
\tempurl}


\bibitem[Deis~Jr and Giroux(1992)]%
        {deis1992determinants}
\bibfield{author}{\bibinfo{person}{Donald~R Deis~Jr} {and} \bibinfo{person}{Gary~A Giroux}.} \bibinfo{year}{1992}\natexlab{}.
\newblock \showarticletitle{Determinants of audit quality in the public sector}.
\newblock \bibinfo{journal}{\emph{Accounting review}} (\bibinfo{year}{1992}), \bibinfo{pages}{462--479}.
\newblock


\bibitem[Feldman et~al\mbox{.}(2015)]%
        {feldman2015certifying}
\bibfield{author}{\bibinfo{person}{Michael Feldman}, \bibinfo{person}{Sorelle~A Friedler}, \bibinfo{person}{John Moeller}, \bibinfo{person}{Carlos Scheidegger}, {and} \bibinfo{person}{Suresh Venkatasubramanian}.} \bibinfo{year}{2015}\natexlab{}.
\newblock \showarticletitle{Certifying and removing disparate impact}. In \bibinfo{booktitle}{\emph{proceedings of the 21th ACM SIGKDD international conference on knowledge discovery and data mining}}. \bibinfo{pages}{259--268}.
\newblock


\bibitem[Gerchick and Watson(2023)]%
        {Gerchick2023}
\bibfield{author}{\bibinfo{person}{Marissa Gerchick} {and} \bibinfo{person}{Brooke Watson}.} \bibinfo{year}{2023}\natexlab{}.
\newblock \showarticletitle{Tracking Automated Employment Decision Tool Bias Audits}.
\newblock  (\bibinfo{year}{2023}).
\newblock
\urldef\tempurl%
\url{https://github.com/aclu-national/tracking-ll144-bias-audits}
\showURL{%
\tempurl}


\bibitem[Goodman and Trehu(2022)]%
        {goodman2022algorithmic}
\bibfield{author}{\bibinfo{person}{Ellen~P Goodman} {and} \bibinfo{person}{Julia Trehu}.} \bibinfo{year}{2022}\natexlab{}.
\newblock \showarticletitle{Algorithmic Auditing: Chasing AI Accountability}.
\newblock \bibinfo{journal}{\emph{Santa Clara High Tech. LJ}}  \bibinfo{volume}{39} (\bibinfo{year}{2022}), \bibinfo{pages}{289}.
\newblock


\bibitem[Groves et~al\mbox{.}(2022)]%
        {Groves2022}
\bibfield{author}{\bibinfo{person}{Lara Groves}, \bibinfo{person}{Jenny Brennan}, \bibinfo{person}{Inioluwa~Deborah Raji}, \bibinfo{person}{Aidan Peppin}, {and} \bibinfo{person}{Strait}.} \bibinfo{year}{2022}\natexlab{}.
\newblock \showarticletitle{Algorithmic impact assessment: a case study in healthcare}.
\newblock  (\bibinfo{year}{2022}).
\newblock
\urldef\tempurl%
\url{https://www.adalovelaceinstitute.org/wp- content/uploads/2022/02/Algorithmic-impact-assessment-a-case-study-in-healthcare.pdf}
\showURL{%
\tempurl}


\bibitem[Indeed(2023)]%
        {Indeed2023}
\bibfield{author}{\bibinfo{person}{Indeed}.} \bibinfo{year}{2023}\natexlab{}.
\newblock \showarticletitle{The Indeed Global AI Survey: Your Guide to the Future of Hiring}.
\newblock  (\bibinfo{year}{2023}).
\newblock
\urldef\tempurl%
\url{https://www.indeed.com/lead/the-indeed-ai-report?hl=en#form}
\showURL{%
\tempurl}


\bibitem[Ivanova(2020)]%
        {Ivanova2020}
\bibfield{author}{\bibinfo{person}{Irina Ivanova}.} \bibinfo{year}{2020}\natexlab{}.
\newblock \showarticletitle{New York City wants to restrict artificial intelligence in hiring}.
\newblock \bibinfo{journal}{\emph{CBS News}} (\bibinfo{year}{2020}).
\newblock
\urldef\tempurl%
\url{https://www.cbsnews.com/news/new-york-city-artificial-intelligence-hiring-restriction/}
\showURL{%
\tempurl}


\bibitem[Jones et~al\mbox{.}(2023)]%
        {Jones2023}
\bibfield{author}{\bibinfo{person}{Elliot Jones}, \bibinfo{person}{Jenny Brennan}, \bibinfo{person}{Connor Dunlop}, {and} \bibinfo{person}{Andrew Strait}.} \bibinfo{year}{2023}\natexlab{}.
\newblock \showarticletitle{Keeping an eye on AI}.
\newblock \bibinfo{journal}{\emph{Ada Lovelace Institute}} (\bibinfo{year}{2023}).
\newblock
\urldef\tempurl%
\url{https://www.adalovelaceinstitute.org/wp-content/uploads/2023/09/ALI_Keeping-an-eye-on-AI-2023.pdf}
\showURL{%
\tempurl}


\bibitem[Kirchner(2023)]%
        {Kirchner2023}
\bibfield{author}{\bibinfo{person}{Lauren Kirchner}.} \bibinfo{year}{2023}\natexlab{}.
\newblock \showarticletitle{New York City Moves to Create Accountability for Algorithms}.
\newblock \bibinfo{journal}{\emph{ProPublica}} (\bibinfo{year}{2023}).
\newblock
\urldef\tempurl%
\url{https://www.propublica.org/article/new-york-city-moves-to-create-accountability-for-algorithms}
\showURL{%
\tempurl}


\bibitem[Koshiyama et~al\mbox{.}(2021)]%
        {koshiyama2021towards}
\bibfield{author}{\bibinfo{person}{Adriano Koshiyama}, \bibinfo{person}{Emre Kazim}, \bibinfo{person}{Philip Treleaven}, \bibinfo{person}{Pete Rai}, \bibinfo{person}{Lukasz Szpruch}, \bibinfo{person}{Giles Pavey}, \bibinfo{person}{Ghazi Ahamat}, \bibinfo{person}{Franziska Leutner}, \bibinfo{person}{Randy Goebel}, \bibinfo{person}{Andrew Knight}, {et~al\mbox{.}}} \bibinfo{year}{2021}\natexlab{}.
\newblock \showarticletitle{Towards algorithm auditing: a survey on managing legal, ethical and technological risks of AI, ML and associated algorithms}.
\newblock  (\bibinfo{year}{2021}).
\newblock


\bibitem[Leslie(2019)]%
        {leslie2019understanding}
\bibfield{author}{\bibinfo{person}{David Leslie}.} \bibinfo{year}{2019}\natexlab{}.
\newblock \showarticletitle{Understanding artificial intelligence ethics and safety}.
\newblock \bibinfo{journal}{\emph{arXiv preprint arXiv:1906.05684}} (\bibinfo{year}{2019}).
\newblock


\bibitem[Lohr(2023)]%
        {Lohr2023}
\bibfield{author}{\bibinfo{person}{Steve Lohr}.} \bibinfo{year}{2023}\natexlab{}.
\newblock \showarticletitle{A Hiring Law Blazes a Path for A.I. Regulation}.
\newblock \bibinfo{journal}{\emph{The New York Times}} (\bibinfo{year}{2023}).
\newblock
\urldef\tempurl%
\url{https://www.nytimes.com/2023/05/25/technology/ai-hiring-law-new-york.html}
\showURL{%
\tempurl}


\bibitem[of~Consumer and Protections(2023a)]%
        {NYC_DCWP2023}
\bibfield{author}{\bibinfo{person}{New York City~Department of Consumer} {and} \bibinfo{person}{Worker Protections}.} \bibinfo{year}{2023}\natexlab{a}.
\newblock \showarticletitle{Automated Employment Decision Tools: Frequently Asked Questions}.
\newblock  (\bibinfo{year}{2023}).
\newblock
\urldef\tempurl%
\url{https://www.nyc.gov/assets/dca/downloads/pdf/about/DCWP-AEDT-FAQ.pdf}
\showURL{%
\tempurl}


\bibitem[of~Consumer and Protections(2023b)]%
        {NYC_AEDT2023}
\bibfield{author}{\bibinfo{person}{New York City~Department of Consumer} {and} \bibinfo{person}{Worker Protections}.} \bibinfo{year}{2023}\natexlab{b}.
\newblock \showarticletitle{Automated Employment Decision Tools (Updated)}.
\newblock  (\bibinfo{year}{2023}).
\newblock
\urldef\tempurl%
\url{https://rules.cityofnewyork.us/rule/automated-employment-decision-tools-updated/}
\showURL{%
\tempurl}


\bibitem[of~Insurance(2023)]%
        {coloradoInsurance2023}
\bibfield{author}{\bibinfo{person}{Colorado~Division of Insurance}.} \bibinfo{year}{2023}\natexlab{}.
\newblock \showarticletitle{Regulation 10-1-1 Governance and Risk Management Framework Requirements for Life Insurers’ Use of External Consumer Data and Information Sources, Algorithms, and Predictive Models}.
\newblock  (\bibinfo{year}{2023}).
\newblock
\urldef\tempurl%
\url{https://drive.google.com/file/d/1dlPKJCDo76iHfJZDopQEhTDCmKbuYnNI/view?usp=embed_facebook}
\showURL{%
\tempurl}


\bibitem[of~Justice(2015)]%
        {DOJcfa}
\bibfield{author}{\bibinfo{person}{US~Department of Justice}.} \bibinfo{year}{2015}\natexlab{}.
\newblock \showarticletitle{9-48.000 - Computer Fraud and Abuse Act}.
\newblock  (\bibinfo{year}{2015}).
\newblock
\urldef\tempurl%
\url{https://www.justice.gov/jm/jm-9-48000-computer-fraud}
\showURL{%
\tempurl}


\bibitem[Office(2020)]%
        {ICO2020}
\bibfield{author}{\bibinfo{person}{Information~Commissioner’s Office}.} \bibinfo{year}{2020}\natexlab{}.
\newblock \showarticletitle{Guidance on the AI auditing framework Draft guidance for consultation. Information Commissioner’s Office}.
\newblock  (\bibinfo{year}{2020}).
\newblock
\urldef\tempurl%
\url{https://ico.org.uk/media/2617219/guidance-on-the-ai-auditing-framework-draft-for-consultation.pdf}
\showURL{%
\tempurl}


\bibitem[Parliament(2022)]%
        {UKonline2022}
\bibfield{author}{\bibinfo{person}{UK Parliament}.} \bibinfo{year}{2022}\natexlab{}.
\newblock \showarticletitle{Online Safety Act 2023}.
\newblock  (\bibinfo{year}{2022}).
\newblock
\urldef\tempurl%
\url{https://bills.parliament.uk/bills/3137}
\showURL{%
\tempurl}


\bibitem[Pattison-Gordon(2023)]%
        {PG2023}
\bibfield{author}{\bibinfo{person}{Jules Pattison-Gordon}.} \bibinfo{year}{2023}\natexlab{}.
\newblock \showarticletitle{Colorado Aims to Prevent AI-Driven Discrimination in Insurance}.
\newblock \bibinfo{journal}{\emph{GovTech}} (\bibinfo{year}{2023}).
\newblock
\urldef\tempurl%
\url{https://www.govtech.com/policy/colorado-aims-to-prevent-ai-driven-discrimination-in-insurance}
\showURL{%
\tempurl}


\bibitem[PCAOB({[n.\,d.]})]%
        {PCAOB}
\bibfield{author}{\bibinfo{person}{PCAOB}.} \bibinfo{year}{[n.\,d.]}\natexlab{}.
\newblock \showarticletitle{Driving improvement in audit quality to protect investors}.
\newblock  (\bibinfo{year}{[n.\,d.]}).
\newblock
\urldef\tempurl%
\url{https://pcaobus.org/}
\showURL{%
\tempurl}


\bibitem[Radiya-Dixit and Neff(2023)]%
        {radiya2023sociotechnical}
\bibfield{author}{\bibinfo{person}{Evani Radiya-Dixit} {and} \bibinfo{person}{Gina Neff}.} \bibinfo{year}{2023}\natexlab{}.
\newblock \showarticletitle{A Sociotechnical Audit: Assessing Police Use of Facial Recognition}. In \bibinfo{booktitle}{\emph{Proceedings of the 2023 ACM Conference on Fairness, Accountability, and Transparency}}. \bibinfo{pages}{1334--1346}.
\newblock


\bibitem[Raghavan et~al\mbox{.}(2020)]%
        {raghavan2020mitigating}
\bibfield{author}{\bibinfo{person}{Manish Raghavan}, \bibinfo{person}{Solon Barocas}, \bibinfo{person}{Jon Kleinberg}, {and} \bibinfo{person}{Karen Levy}.} \bibinfo{year}{2020}\natexlab{}.
\newblock \showarticletitle{Mitigating bias in algorithmic hiring: Evaluating claims and practices}. In \bibinfo{booktitle}{\emph{Proceedings of the 2020 conference on fairness, accountability, and transparency}}. \bibinfo{pages}{469--481}.
\newblock


\bibitem[Raji and Buolamwini(2019)]%
        {raji2019actionable}
\bibfield{author}{\bibinfo{person}{Inioluwa~Deborah Raji} {and} \bibinfo{person}{Joy Buolamwini}.} \bibinfo{year}{2019}\natexlab{}.
\newblock \showarticletitle{Actionable auditing: Investigating the impact of publicly naming biased performance results of commercial ai products}. In \bibinfo{booktitle}{\emph{Proceedings of the 2019 AAAI/ACM Conference on AI, Ethics, and Society}}. \bibinfo{pages}{429--435}.
\newblock


\bibitem[Raji et~al\mbox{.}(2022)]%
        {raji2022outsider}
\bibfield{author}{\bibinfo{person}{Inioluwa~Deborah Raji}, \bibinfo{person}{Peggy Xu}, \bibinfo{person}{Colleen Honigsberg}, {and} \bibinfo{person}{Daniel Ho}.} \bibinfo{year}{2022}\natexlab{}.
\newblock \showarticletitle{Outsider oversight: Designing a third party audit ecosystem for ai governance}. In \bibinfo{booktitle}{\emph{Proceedings of the 2022 AAAI/ACM Conference on AI, Ethics, and Society}}. \bibinfo{pages}{557--571}.
\newblock


\bibitem[Rekenkamer(2022)]%
        {Rekenkamer2022}
\bibfield{author}{\bibinfo{person}{Algemene Rekenkamer}.} \bibinfo{year}{2022}\natexlab{}.
\newblock \showarticletitle{An Audit of 9 Algorithms used by the Dutch Government - Report - Netherlands Court of Audit}.
\newblock  (\bibinfo{year}{2022}).
\newblock
\urldef\tempurl%
\url{https://english.rekenkamer.nl/publications/reports/2022/05/18/an-audit-of-9-algorithms-used-by-the-dutch-government}
\showURL{%
\tempurl}


\bibitem[Rep.~Clarke(2022)]%
        {Clarke2022}
\bibfield{author}{\bibinfo{person}{Yvette D. [D-NY-9] Rep.~Clarke}.} \bibinfo{year}{2022}\natexlab{}.
\newblock \showarticletitle{Algorithmic Accountability Act of 2022}.
\newblock  (\bibinfo{year}{2022}).
\newblock
\urldef\tempurl%
\url{https://www.congress.gov/bill/117th-congress/house-bill/6580/text}
\showURL{%
\tempurl}


\bibitem[Selbst(2017)]%
        {selbst2017disparate}
\bibfield{author}{\bibinfo{person}{Andrew~D Selbst}.} \bibinfo{year}{2017}\natexlab{}.
\newblock \showarticletitle{Disparate impact in big data policing}.
\newblock \bibinfo{journal}{\emph{Ga. L. Rev.}}  \bibinfo{volume}{52} (\bibinfo{year}{2017}), \bibinfo{pages}{109}.
\newblock


\bibitem[Selbst et~al\mbox{.}(2019)]%
        {selbst2019fairness}
\bibfield{author}{\bibinfo{person}{Andrew~D Selbst}, \bibinfo{person}{Danah Boyd}, \bibinfo{person}{Sorelle~A Friedler}, \bibinfo{person}{Suresh Venkatasubramanian}, {and} \bibinfo{person}{Janet Vertesi}.} \bibinfo{year}{2019}\natexlab{}.
\newblock \showarticletitle{Fairness and abstraction in sociotechnical systems}. In \bibinfo{booktitle}{\emph{Proceedings of the conference on fairness, accountability, and transparency}}. \bibinfo{pages}{59--68}.
\newblock


\bibitem[SHRM(2022)]%
        {SHRM2022}
\bibfield{author}{\bibinfo{person}{SHRM}.} \bibinfo{year}{2022}\natexlab{}.
\newblock \showarticletitle{Automation and AI in HR}.
\newblock  (\bibinfo{year}{2022}).
\newblock
\urldef\tempurl%
\url{https://advocacy.shrm.org/SHRM-2022-Automation-AI- Research.pdf?_ga=2.112869508.1029738808.1666019592-61357574.1655121608}
\showURL{%
\tempurl}


\bibitem[Tepalagul and Lin(2015)]%
        {tepalagul2015auditor}
\bibfield{author}{\bibinfo{person}{Nopmanee Tepalagul} {and} \bibinfo{person}{Ling Lin}.} \bibinfo{year}{2015}\natexlab{}.
\newblock \showarticletitle{Auditor independence and audit quality: A literature review}.
\newblock \bibinfo{journal}{\emph{Journal of Accounting, Auditing \& Finance}} \bibinfo{volume}{30}, \bibinfo{number}{1} (\bibinfo{year}{2015}), \bibinfo{pages}{101--121}.
\newblock


\bibitem[Watkins et~al\mbox{.}(2022)]%
        {watkins2022four}
\bibfield{author}{\bibinfo{person}{Elizabeth~Anne Watkins}, \bibinfo{person}{Michael McKenna}, {and} \bibinfo{person}{Jiahao Chen}.} \bibinfo{year}{2022}\natexlab{}.
\newblock \showarticletitle{The four-fifths rule is not disparate impact: a woeful tale of epistemic trespassing in algorithmic fairness}.
\newblock \bibinfo{journal}{\emph{arXiv preprint arXiv:2202.09519}} (\bibinfo{year}{2022}).
\newblock


\end{thebibliography}
\bibliographystyle{ACM-Reference-Format}

\appendix

\section{Appendices}
\FloatBarrier\subsection{Example audit report, prepared by ConductorAI for National Broadcasting Company (NBC)}

\begin{figure*}[]
  \centering
  \includegraphics[width=\linewidth]{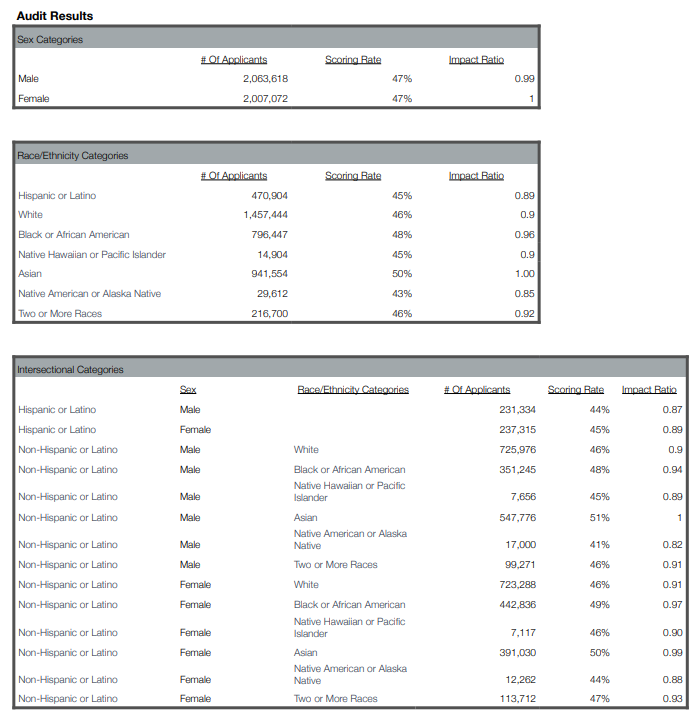}
   \caption{Example audit report, prepared by ConductorAI for NBC}
  \label{fig:fig3}
\end{figure*}

\FloatBarrier\subsection{Table 2: Participant ID}
\begin{figure*}[]
  \centering
  \includegraphics[width=\linewidth]{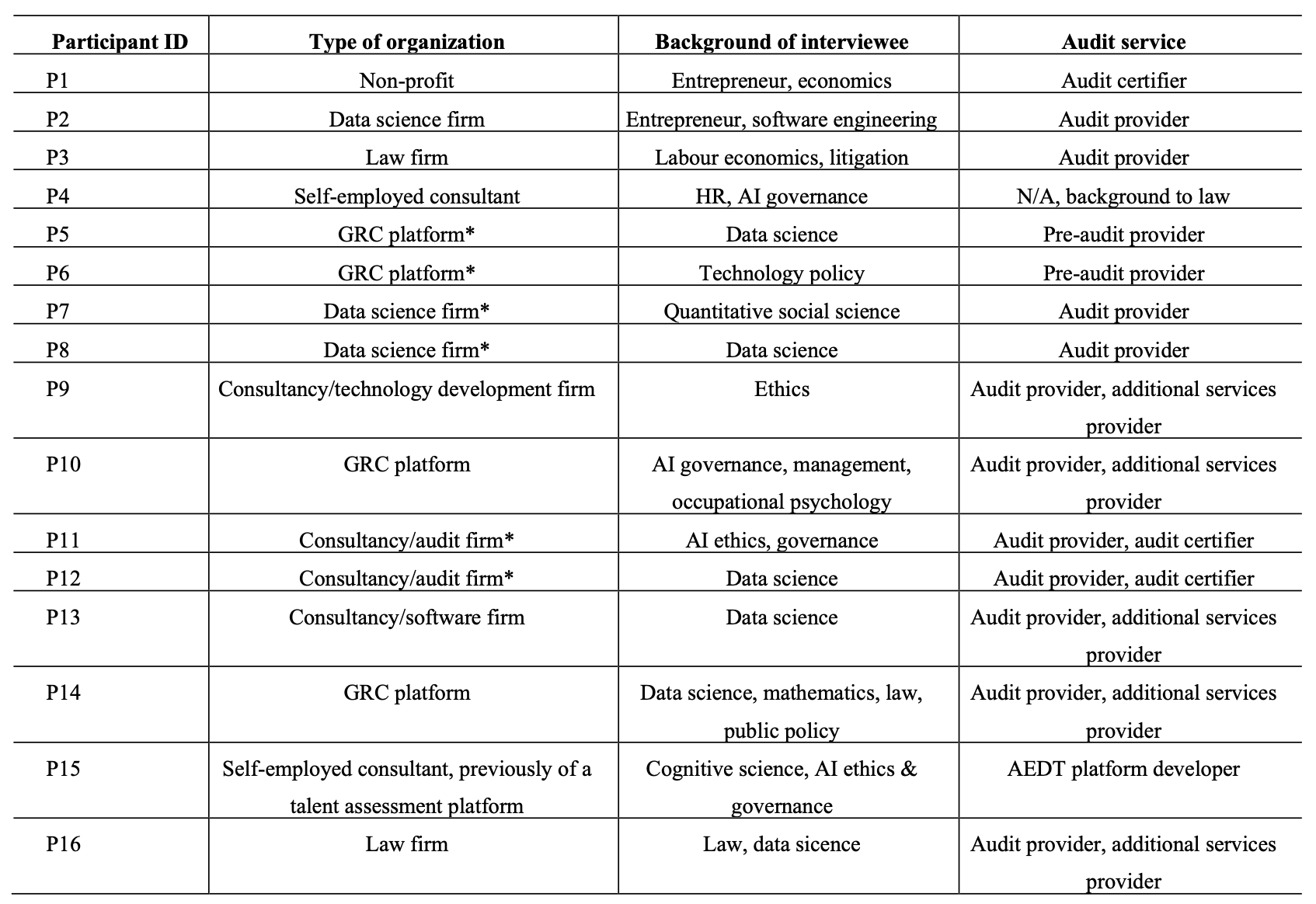}
   \caption*{Table 2: Participant ID}
   \caption*{Entries marked with an asterisk represent instances where we interviewed two representatives from the same organisation.}
  \label{fig:tab2}
\end{figure*}

\end{document}